# Cyanine-Conjugated Gold Nanospheres for Near-infrared Fluorescence

*Neelima Chacko[1], Menachem Motiei[2], Nitzan Dar[1] and Rinat Ankri\*[1]*

1. Department of Physics, Faculty of Natural Science, Ariel University, Ariel 40700, Israel

2. Faculty of Engineering & The Institute of Nanotechnology and Advanced Materials, Bar-Ilan University, Ramat Gan 5290002, Israel

*Corresponding author: rinatsel@ariel.ac.il

**Abstract**

Near-infrared (NIR) fluorescence imaging offers improved spatial precision by reducing light scattering and absorption in tissue. Despite this key advantage, the NIR region is limited by the availability of fluorophores, most of which exhibit relatively low quantum yield (QY). In this study, gold nanospheres (AuNSs) with absorption peaks in the visible range (400–700 nm) were used to enhance the fluorescence intensity of the cyanine NIR fluorophore IRdye 800 in the first NIR window (NIR-I, 700–900 nm) of the electromagnetic spectrum. AuNSs with diameters ranging from 5 to 25 nm were chosen to investigate the impact of a nanoparticle size on fluorescence enhancement, functionalized with polyethylene glycol (PEG) of varying molecular weights (1 kDa, 2 kDa, 5 kDa, and 7.5 kDa) to optimize the distance between the fluorophore and the nanoparticle surface. Theoretical analyses using finite-difference time-domain (FDTD) simulations and experimental comparisons with non-metallic nanoparticles were performed to identify the factors contributing to the enhancement of fluorescence. PEGylated AuNSs conjugated with IRdye 800 (AuNDs) exhibited decreased photoisomerization, resulting in increased fluorescence intensity and altered fluorescence lifetimes (FLTs). The observed enhancement in the fluorescence intensity of the AuNDs was attributed to three primary mechanisms: metal-enhanced fluorescence, altered radiative decay rates, and steric stabilization. Among these three mechanisms, two are attributed to the tail-end absorption spectral overlap of the AuNSs with IRdye 800. This study highlights the potential of AuNSs for improving NIR-I fluorescence imaging and opens up new possibilities for applications in biomedical research.

**Keywords:** NIR-I region, Gold nanoparticles, Fluorescence enhancement, Surface plasmon resonance, Steric stabilization.

**Table of contents**

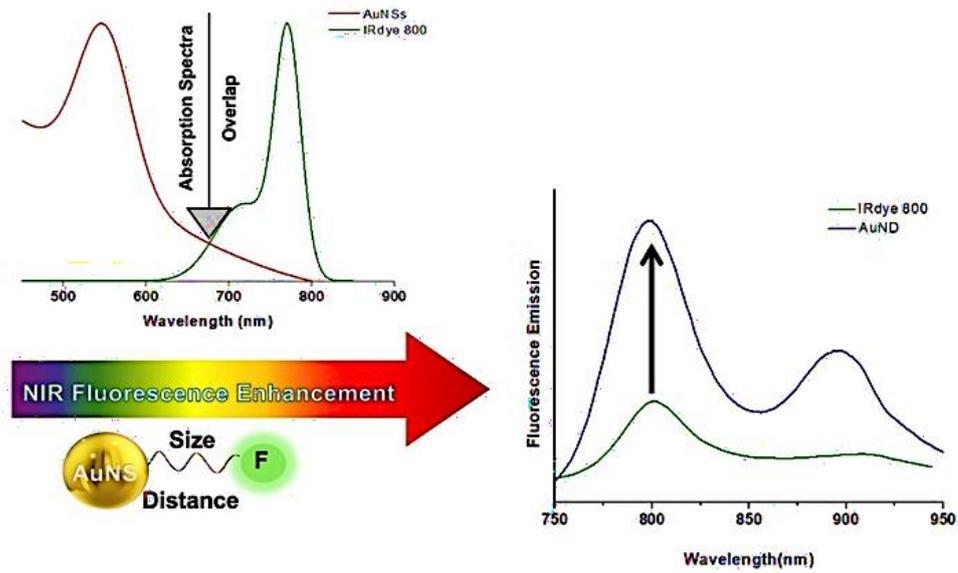

**Introduction**

Gold nanoparticles (AuNPs) have garnered significant attention in recent years for their potential applications in biomedical imaging, largely due to their distinctive physicochemical properties.[1] They exhibit strong absorption and scattering properties[2,3], excellent biocompatibility[1], simple synthesis[4,5], and facile surface functionalization[6,7], making them highly advantageous for biological imaging applications. Gold nanospheres (AuNSs) are predominantly used for biological imaging due to their unique localized surface plasmon resonance (LSPR) properties[8,9]. Despite their relatively easy fabrication and strong photophysical properties, AuNSs have been relatively less studied in fluorescence imaging applications. This is primarily due to their surface plasmon resonance (SPR) characteristics in the visible region, which can result in lower tissue penetration for bioimaging[10]. In this study, we demonstrate that PEGylated AuNSs can be employed for fluorescence imaging in the first near-infrared (NIR-I, 760-900 nm) region by their conjugation to a NIR fluorophore, IRdye 800 NHS ester, which is a cyanine dye currently in phase II clinical trials[11].

The IR800 NHS ester dye, an amine-reactive form of the fluorophore IRdye 800, is commonly used in bio-imaging to generate stable fluorescence signals[11]. It efficiently binds to nucleophiles, particularly to the epsilon-amino groups of solvent-accessible lysine residues on various substrates such as antibodies, proteins, peptides, enzymes, and other small molecules[12]. Despite its advantages, the IR800 NHS ester dye has a low quantum yield ($QY = 0.03\%$)[13], which restricts its effectiveness for in vitro and in vivo studies. Extensive research has concentrated on enhancing its fluorescence intensity and biocompatibility, primarily through chemical stabilization and modifications, which are often more challenging to implement[14,15]. Recently, we proposed that the conjugation of PEG-functionalized AuNSs with IRdye 800 (referred to as Au nanodyes, or AuNDs) can significantly enhance the fluorescence of the dye without requiring any chemical modifications[16]. In this work, we build upon our previous findings and conduct a detailed exploration of the key factors contributing to the enhanced fluorescence of IRdye 800 when conjugated to AuNSs.

This study presents a comprehensive analysis of the fluorescence properties of self-synthesized AuNDs, demonstrating that AuNSs, with a tailing absorption at >700 nm, can be utilized for NIR-I fluorescence imaging by conjugating PEGylated AuNSs to IRdye 800. The fluorescence enhancement of IRdye 800 by AuNSs is attributed to three key factors: (i) the metal-enhanced fluorescence (MEF) effect, (ii) alterations in the dye's radiative decay rate, and (iii) the steric stabilization of the NIR dye due to AuNSs. These factors are thoroughly investigated through theoretical approaches, including finite-difference time-domain (FDTD) simulations to assess the electric field of AuNSs in the NIR-I spectrum, complemented by experimental studies. The study also explores the impact of varying the diameters of AuNSs (5, 10, 15, 20, and 25 nm) on the fluorescence of the dye, as well as the effects of adjusting the distance between the dye and the metallic nanoparticle using polyethylene glycol (PEG) chains with different molecular weights (1 kDa, 2 kDa, 5 kDa, and 7.5 kDa). Additionally, the study evaluates the steric stabilization effects of nanoparticles on fluorescence enhancement by analyzing the fluorescence of IRdye 800 in conjunction with non-metallic ZnO nanoparticles.

**Materials and Methods**

The AuNS of diameters 5,10,15,20, and 25 were purchased from nanopartz (Colorado, USA). Their TEM images, absorption and emission spectra are presented in the Supporting Information (Figure S1 and Figure S2). The SH-mPEG (an average molecular weight (M.W.) of 6 kDa) and SH-PEG-NH$_2$ (average M.W. of 1kDa, 2kDa, 5kDa, and 7.5kDa) were purchased from Sigma Aldrich (Israel) and Biopeg pharma (USA). The IRDye 800 used in the experiments was purchased from LI-COR Biosciences (USA).

*AuNSs PEG coating*

Poly (ethylene glycol) methyl ether thiol (mPEG-SH) of average M.W. 6 kDa and thiol poly(ethylene glycol) amine (NH$_2$-PEG-SH) with average molecular weights of 1 kDa, 2 kDa, 5 kDa, and 7.5 kDa were dissolved separately in ultra-pure Milli-Q water to create stock solutions with concentrations of 35 mg/mL.NH$_2$-PEG-SH serves as a spacer and linker between the surface of the AuNS and the dye. On the other hand, mPEG-SH provides hydrophilicity and stability to the AuNS, while also regulating the amount of dye on the AuNS surface. The AuNS were PEGylated using a mixture consisting of 85% mPEG-SH and 15% NH$_2$-PEG-SH, ensuring complete coverage of the AuNS with PEG chains, achieved by determining the available surface area of the different diameter AuNS for PEGylation. Following the addition of the required volume of PEG solutions to the AuNS, the mixture was continuously stirred at room temperature for two hours, followed by two washes at 14,000 rpm at room temperature for 20 minutes.

*IRdye 800 conjugation to AuNSs*

The IRdye 800 (M.W. 1165.10g/mol) was dissolved in PBS of pH 7 to prepare a 1 mM stock solution. Then 11 µL (1mM) of dye is added to the PEGylated AuNS and stirred continuously for two hours at room temperature in a dark environment at pH 7. The NHS ester-activated moiety of the dye, possessing amine reactivity, enables its conjugation to the amine terminal of the $NH_2$-PEG-SH-coated AuNS (Figure 1(a)) and the variation in the dynamic light scattering results for 25 nm AuNSs conjugated to various PEG chain lengths attached to the dye is shown in Figure 1(b). Given that each sample had a unique dye loading, it was required to calculate the expected fluorescence emission for each. This involved creating calibration curves that correlated fluorescence peak intensity with free dye concentration (Supporting Information, Figure S3). The fluorescence emission of the supernatants was then used to quantify the concentration of residual free dye in each sample supernatant following incubation with the AuNSs and subsequent centrifugation. Consequently, the dye loading on the AuNSs was calculated.

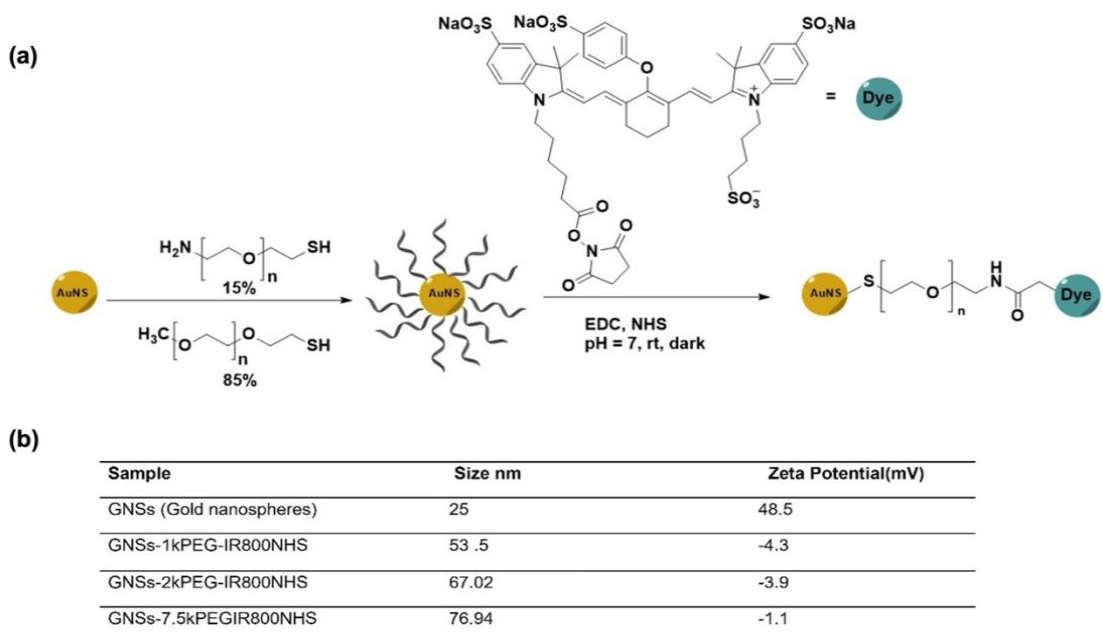

| Sample | Size nm | Zeta Potential(mV) |
|---|---|---|
| GNSs (Gold nanospheres) | 25 | 48.5 |
| GNSs-1kPEG-IR800NHS | 53.5 | -4.3 |
| GNSs-2kPEG-IR800NHS | 67.02 | -3.9 |
| GNSs-7.5kPEGIR800NHS | 76.94 | -1.1 |

**Figure 1: Synthesis and DLS characterization of AuNDs** **(a)** Schematic representation of AuND synthesis via PEGylation of AuNSs followed by conjugation of IRDye 800 to the PEGylated AuNSs. **(b)** The tabulated data summarizes the mean hydrodynamic diameters and zeta potential

measurements (at 25 °C) of PEGylated 25 nm AuNSs (a representative sample) conjugated to the dye.

*Synthesis of ZnO nanoparticles*

The ZnO nanoparticles were synthesized following a modified version of a previously reported method[17]. Specifically, 1.367 mg of zinc acetate and 0.786 mg of oxalic acid were dissolved in 15 mL of water and stirred continuously for two hours at a temperature of 27 °C. Excess water was then removed using a rotary evaporator at 45 °C. The resulting white solid powder was washed with methanol and filtered at least twice, followed by drying in an oven at 60 °C overnight. The dried product was then grinding thoroughly for 30 minutes at room temperature using an agate mortar, forming $ZnC_2O_4 \cdot 2H_2O$ nanoparticles. Finally, the ZnO nanoparticles were obtained by calcining the $ZnC_2O_4 \cdot 2H_2O$ nanoparticles at 400 °C for 3 hours. The synthesized ZnO nanoparticles are PEGylated and conjugated to the dye as described in the above procedure.

*Steady-state fluorescence measurements*

Absorbance and emission spectra were measured with the Fluorolog-Quanta Master (Horiba scientific, Japan). The data were analyzed using the FelixFL software (version 1.0.33.0, Horiba Scientific, Japan). Quartz cuvettes with an optical path length of 1 cm were used for both absorbance and fluorescence measurements. For the absorption and emission measurements, the samples were excited with a 75 W Xenon lamp at a slit width of 10 nm and an integration time of 1 s with a step size of 0.1 nm in the wavelength.

*Time-correlated single photon counting (TCSPC) lifetime measurements*

The lifetime measurements were performed using the Fluorolog-Quanta Master (Horiba scientific, Japan), using a delta diode of 830 ± 10 nm (a peak wavelength of 819 ± 10 nm), with a narrow 50 picoseconds (ps) pulse width, a 0.6 mW average power, and a 100 MHz repetition rate. The samples were placed in a 1 cm pathlength absorbance plastic cuvette. Instrument response function (IRF) was measured using LUDOX® AS-30 colloidal silica (Sigma Aldrich, Israel) and an excitation wavelength of 805 nm through a 35nm slit width. The fluorescence decay curves were analyzed using the FelixFL decay analysis software (version 1.0.33.0, Horiba Scientific) based on a multiexponential model which involves an iterative reconvolution process. The quality of the fits was evaluated using the reduced $\chi^2$ value.

*Wide field FLI setup*

The excitation source was a fiber-coupled pulsed laser with a wavelength of 779 nm, 20 MHz repetition rate, and a pulse width of ~ 70 ps (VisIR-780, PicoQuant). Wide-field illumination was achieved with a 10x beam expander (GBE10-B, Thorlabs). The dye solutions were placed between two glass slides separated by a silicon insulator film (Merck, Israel). The emitted fluorescence was recorded with a NIR objective lens (5018- SW, Computar, TX) and imaged with the SPAD512S camera (PiImaging, Switzerland). The time-triggered in-pixel architecture enables time-resolved photon counting at a maximum rate of 97 kfps (1-bit frames). The photon detection efficiency of the detector was ~13% at 800 nm with a fill factor of 10.5% and a dark count rate with a median value of 7.5 Hz/pixel. Overlapping gate images (G = 117, gate spacing g_s = 428 ps) were used, with a gate image acquisition time of 50 ms in each experiment. Excitation power at focal plane: 5 mW, total integration time: 25 s.

*Phasor based analyses*

Phasor analyzes of the wide-field time-gated FLI data was performed using the free software *Alligator* [18] for phasor fluorescence lifetime (FLT) analyses. In the basis of these programs, the phasor ($g_{i,j}$, $s_{i,j}$) of each pixel of coordinate (*i, j*) in the image is calculated according to:

(1)
$$g_{i,j} = \frac{\sum_{k=1}^{N} F_{i,j}(t_k) \cos(2\pi f t_k)}{\sum_{k=1}^{N} F_{i,j}(t_k)}$$
$$s_{i,j} = \frac{\sum_{k=1}^{N} F_{i,j}(t_k) \sin(2\pi f t_k)}{\sum_{k=1}^{N} F_{i,j}(t_k)}$$

where *f* is the phasor harmonic (equal to the laser repetition rate = 1/*T*), k = 1 … N is the gate number and $F_{i,j}(t_k)$ is the k[th] gate image value at pixel (i, j). When computing region of interest (ROI) phasor values, the $F_{i,j}(t_k)$ in Eq. [1] are replaced by the sum of all $F_{i,j}(t_k)$ in the ROI. The FLT was calculated from:

(2)
$$\tau = \frac{1}{2\pi f} \frac{s}{g}$$

When the mean FLT was given by the mean of the phase lifetime histogram[19].

*Simulation*

The finite-difference time-domain (FDTD) method was used to perform numerical simulations using FDTD solutions software (Lumerical Solutions, 2022). Au nanospheres with diameters of 15, 20, and 25 nm were simulated, along with a point dipole representing IRdye 800, with a wavelength of 800 ± 10 nm and a refractive index of 1.33. The dipole was oriented in the Z-direction and positioned at various approximate distances from PEG chains; 6 nm, 12 nm, and 30 nm, corresponding to 1 kDa, 2 kDa, and 5 kDa PEG chains, respectively. A three-dimensional nonuniform meshing technique was employed, with a grid size of 0.6 nm. The Palik model for Au nanoparticles was used in the simulation[20]. Perfectly matched layer (PML) absorption boundary conditions, along with symmetric boundary conditions, were applied to reduce memory requirements and computational time.

*Statistics*

All data were represented using GraphPad Prism 9.2.0.and origin pro-8. All bar graphs represent ± SD data.

**Results and discussion**

*Photophysical study of the Au nanodyes (AuNDs)*

Figure 2(a) shows the absorption spectra of IRdye 800 and 25 nm AuNSs in PBS, resulting in absorption peaks at 772 nm and 522 nm, respectively. The absorption spectrum of the dye overlaps with the tail end of the AuNS absorption spectrum, forming the basis for the fluorescence enhancement of the Au nanodyes, as will be discussed later. Figure 2(b) shows the fluorescence emission spectra of AuNSs and IRdye 800 and Figure 2(c) the FLT of IRdye 800 and AuNDs, as was measured by the TCSPC set up, as described above, resulting with a lifetime of 0.67 ± 0.01 ns and 0.19 ± 0.01 ns, respectively.

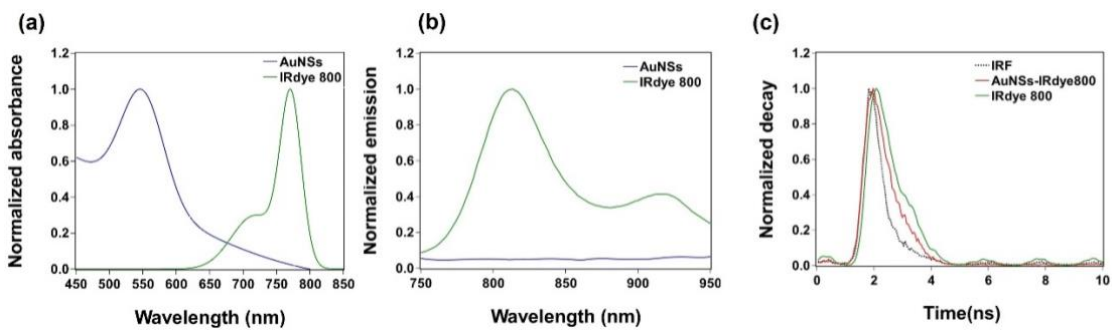

**Figure 2: Photophysical characterization of IRdye 800 and AuNS (25 nm). (a)** Absorption spectra of IRdye 800 (in PBS, $\lambda_{max}$ 772 nm) and AuNS ($\lambda_{max}$ 522 nm). **(b)** The fluorescence emission spectra of AuNSs and IRdye 800 (in PBS, $\lambda_{max}$ 805 nm). **(c)** The fluorescence decay curve of IRdye 800 and 25 nm PEGylated AuNSs conjugated to IRdye 800 (AuND), both in PBS solution. Black line: instrumental response function (IRF). Red line: the fitted dye's emission curve. Green line: the fitted AuND's emission curve.

The AuNSs, with diameters ranging from 5 to 25 nm, were functionalized using PEG chains of different molecular weights. This size range was selected for its relatively enhanced in vitro uptake as well as lower cytotoxicity across various cell types[21–24]. Thus, for example, a recent study by Żurawski et.al. showed that gold nanospheres (with a specific size of 15 nm) exhibit lower toxicity compared to gold nanorods with dimensions 22 × 50 nm and 20 × 40 nm in mammalian cells [25]. We demonstrate the biocompatibility of our synthesized AuNDs in both cancer[16] cells and normal cells through in vitro studies (Supporting Information, Figure S4). The PEG chains act as both spacers and linkers between the dye and AuNSs. By varying the PEG chain length, we could assess the impact of the AuNSs's LSPR on the dye fluorescence at different distances.

Figure 3 shows the absorption spectra of PEGylated AuNSs of 5 nm, 10 nm, 15 nm, 20 nm, and 25 nm in diameter, conjugated to IRdye 800 with the different molecular weight PEG chains: 1 kDa, 2 kDa, 5 kDa, and 7.5 kDa. The AuND are denoted as AuND(i)j; in this notation, "i" represents the radius of the nanospheres, which ranges from 5 nm to 25 nm in increments of 5 nm, and "j" represents the average molecular weight of the PEG chain, which can be 1 kDa, 2 kDa, 5 kDa or 7.5 kDa. For example, an AuND with an AuNS of radius 5 nm conjugated to a 1 kDa PEG chain bound to IRdye 800 is referred to as AuND(5)1. In Figures 3 (a)-(d), the first peak (520-540 nm) corresponds to the LSPR of the gold nanospheres, while the second peak (700-800 nm) arises from the conjugation of these nanoparticles with the NIR dye [25]. AuNDs with 1 kDa and 2 kDa PEGylated 5 nm AuNS have an absorption peak at 768 nm, while 10, 15, 20, and 25 nm AuNS have an absorption peak at 772 nm. 5 kDa PEGylated AuNDs of 10,15 nm have an absorption peak at 784 nm while 20 and 25 nm have absorption peaks at 778 nm and 787 nm, respectively. The absorption peaks of 7.5 kDa PEGylated AuNDs of 10,15,20, and 25 nm are at 773 nm. The absorption curves of all synthesized AuNDs, featuring two distinct peaks, demonstrate a good conjugation between the AuNSs and IRdye 800, with one peak corresponding to the absorption of the AuNSs (between 500-600 nm) and the other to the dye (between 750-800 nm).

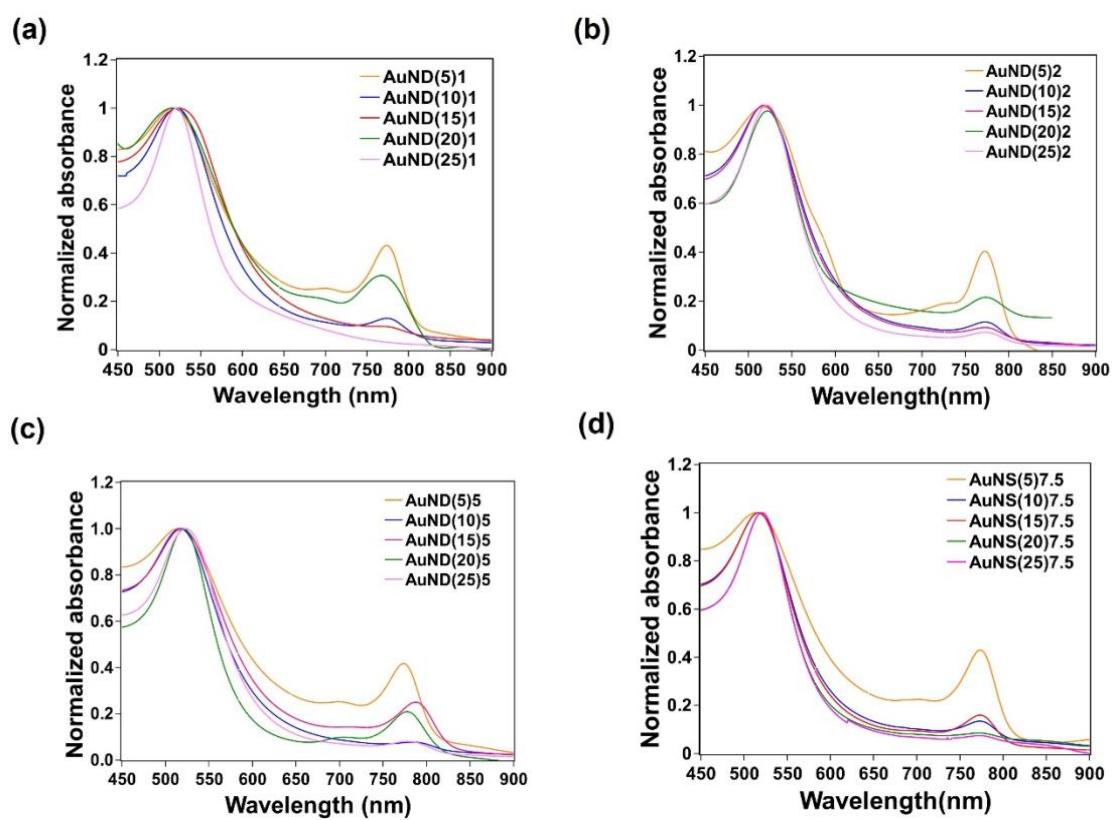

**Figure 3: Normalized absorption spectra of 5 nm,10 nm,15 nm,20 nm, and 25 nm AuNDs**. **(a)** The AuNSs are attached to the dye via 1 kDa PEG chains. **(b)** The AuNSs are attached to the dye via 2 kDa PEG chains. **(c)** The AuNSs are attached to dye via 5 kDa PEG chains. **(d)** The AuNSs are attached to dye via 7.5 kDa PEG chains.

AuNDs with the different molecular weight PEG chains; 1 kDa, 2 kDa, and 5 kDa, and different NPs diameters presented emission peaks ranging from 790 to 816 nm (Figure 4(b)-(f)). It indicates that the presence of the nanoparticles does not significantly affect the dye's emission peak wavelength (Supporting Information, Table S1), as the peak emission wavelength of PEGylated IRdye 800 is at 809 nm (Figure 4(a)). The fluorescence emission peak small shift observed between larger AuNSs conjugated dye (25 nm, ~810 nm) and smaller ones (5 nm, ~799 nm) can be attributed to the LSPR effect of the AuNS; as the dye binds to the PEG chain via an amide bond, it potentially alters its microenvironment and refractive` index compared to its unbound state. LSPR effect is highly sensitive to the refractive index of the surrounding medium, which influences the interaction between light and the nanoparticles. This sensitivity is particularly pronounced for AuNSs, and the extent of this effect is strongly dependent on the size of the nanoparticles. As the size of AuNSs increases, the LSPR effect becomes more pronounced due to the larger surface area available for light interaction, resulting in a shift in the emission peak and enhanced fluorescence for larger nanoparticles conjugated to a dye. The enhanced electromagnetic field generated by LSPR in larger AuNSs increases the radiative decay rate of the dye, thereby improving fluorescence intensity. Consequently, AuNSs with larger diameters are more effective in fluorescence enhancement when conjugated with a dye[26].

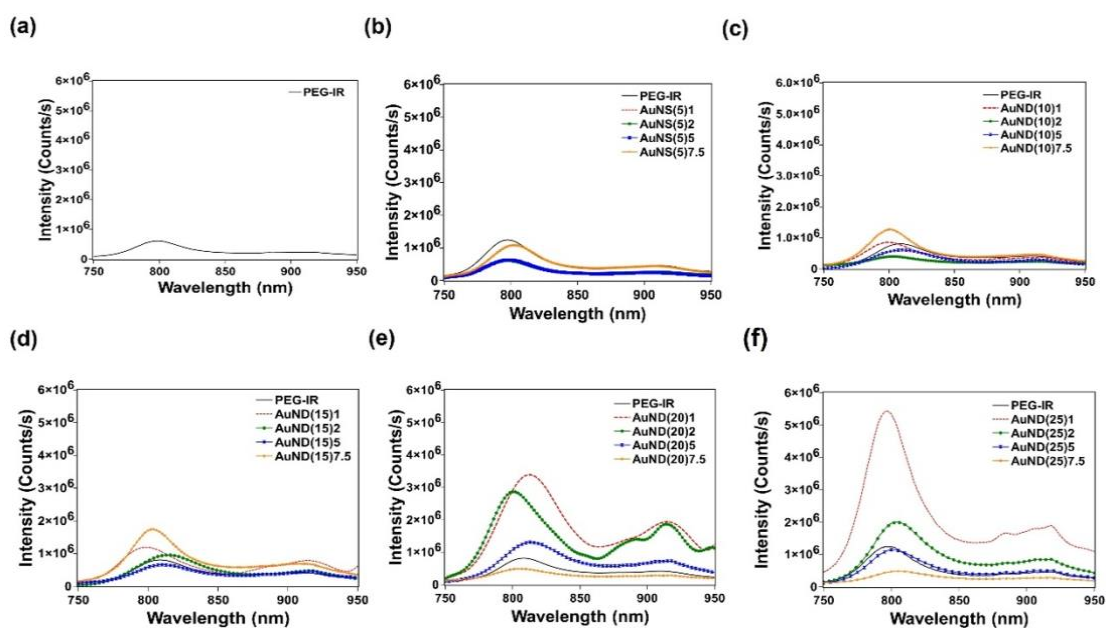

**Figure 4: Fluorescence emission spectra of AuNDs of different diameters conjugated with 1 kDa, 2 kDa, 5 kDa and 7.5 kDa PEG chains.** (a) PEG conjugated IRdye 800 in PBS (pH 7), showing an emission peak at 809 nm. (b) 5 nm AuNDs. (c) 10 nm AuNDs. (d) 15 nm AuNDs. (e) 20 nm AuNDs. (f) 25 nm AuNDs. The fluorescence intensities of AuNDs are normalized according to the conjugated dye concentration using a calibration curve (Supporting Information, Figure S3). The IRdye 800 emission is measured in the same microenvironment as the AuNDs.

*The interplay between fluorescence enhancement, AuNS size, and PEG length*

In Figure 5, we summarize the emission intensity values of the different synthesized AuNDs. Figure 5 (a)-(d) emphasizes the effect of the AuNSs size on the fluorescence emission, while Figure 5 (e-i) demonstrates the effect of different molecular weight PEG chains on the AuNDs emission intensity. Larger-sized AuNDs, specifically those with diameters of 20 nm and 25 nm, resulted in higher fluorescence emission of the dye compared to smaller AuNDs. This reduced effect in smaller AuNSs (5-15 nm) is attributed to the decreased scattering cross-section and lower electric field intensity, which influence the LSPR of the particle[27-29]. Interestingly, a higher fluorescence enhancement was observed in 5 nm AuNSs PEGylated with 1 kDa and 7.5 kDa PEG chains conjugated to dye, compared to 10 nm AuNSs conjugated with the same PEG chains (Figure 5(e) and 5(f)). It can be explained by the fact that small nanoparticles possess high energy levels due to their tendency to aggregate, leading to LSPR and increased light absorption.[30]

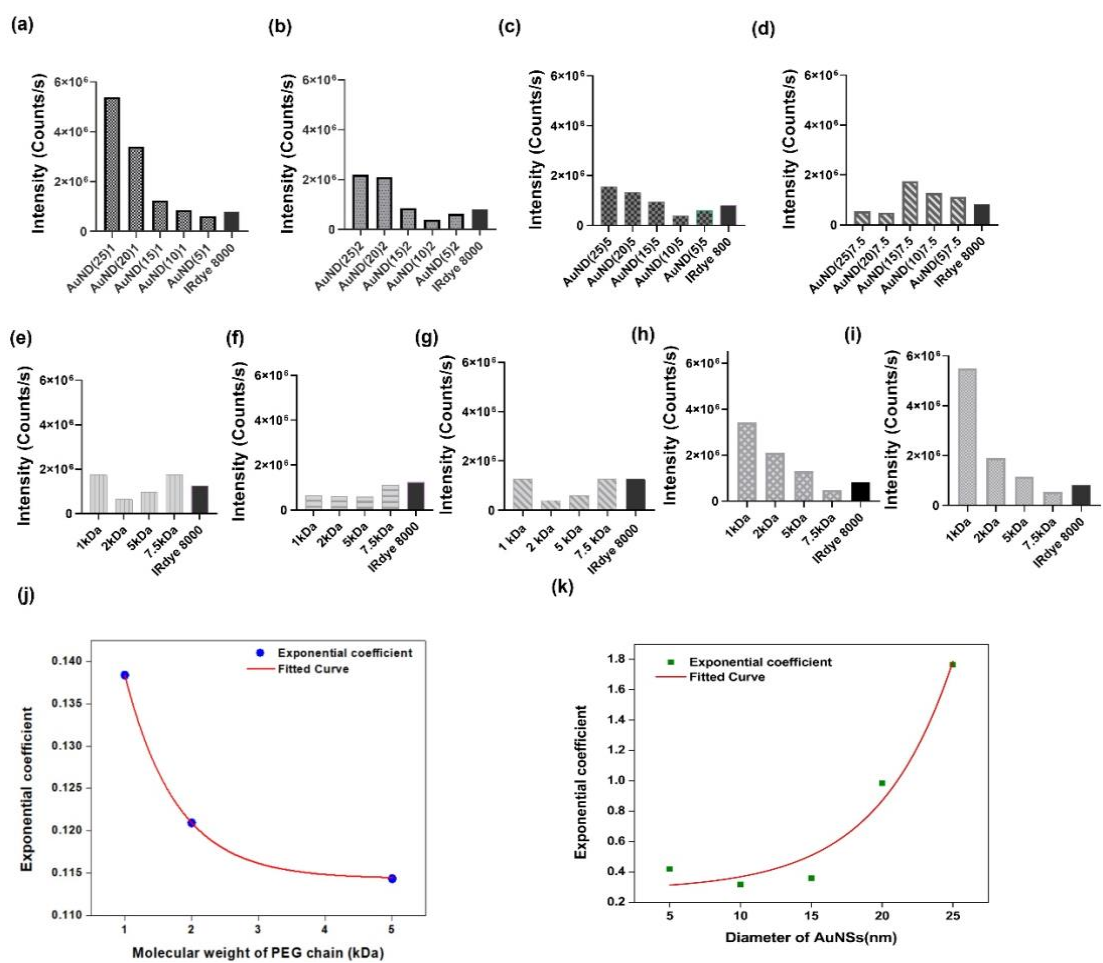

**Figure 5: Fluorescence intensity of AuNDs as a function of PEG chain length and AuNSs diameter. (a-d)** Fluorescence intensity of the AuNDs as function of the molecular weight of the PEG chains. **(e-i)** Fluorescence intensity of the AuNDs as function of the AuNSs diameter. **(j)** The exponential coefficient obtained from the fluorescence intensity of AuNDs conjugated to 1 kDa, 2 kDa, and 5 kDa PEG chains to different diameter AuNS (histograms presented in Figure 5(a-d).) **(k)** The exponential coefficient obtained from the fluorescence intensity of AuNDs of 5,10,15,20 and 25 nm AuNS with different molecular weight PEG chains (1,2, and 5 kDa) (histograms presented in Figure 5(e-i)). The exponential function for Figure 5(j) is: $y = A_1 \left[ e^{-\frac{x}{t}} \right] + y_0$, where the |t| value is 0.773, while the exponential function describing the relationship between the diameter and fluorescence intensity gave the best fit to the following equation: $y = A_1 \left[ e^{\frac{x}{t}} \right] + y_0$, with a |t| value of 5.350.

Figures 5(j) and 5 (k) show the exponential coefficients extracted from the histograms in Figures 5(a-d) and 5(e-i), as a function of the PEG chain length and the NPs size, respectively. The 7.5 kDa PEGylated AuNDs results are excluded from the exponential decay function because of the high flexibility of the long PEG chain, which leads to an unexpected length, resulted in a non-specific emission behaviour. Figure 5(j) illustrates that as the length of the PEG increases, the relationship between the fluorescence intensity and PEG length diminishes, regardless of the nanoparticle size. In contrast, Figure 5(k) demonstrates that as the size of the Au nanospheres increases, the correlation between fluorescence emission and the size of the AuNSs also intensifies. By comparing the exponential factors of the two decay processes, the distance between the fluorophore and the nanoparticle plays a more significant role in fluorescence enhancement than the nanoparticle's diameter.

According to previous reports, the fluorescence enhancement of the NIR dye upon conjugation with AuNSs can be attributed to the following factors; (i) the enhanced electric field near the dye, resulting from the overlap of the tail end of the AuNS' LSPR with the NIR dye's absorption, (ii) the modification in the dye's radiative decay rate, and (iii) the steric stabilization of the dye provided by its conjugation with the functionalized AuNSs. In the following, we investigate the influences of the aforementioned factors on the fluorescence imaging of AuNDs in the NIR region.

*Simulated electric field distribution of AuNDs*

Simulations of the electric field distributions of AuNS with varying diameters and distances from the NIR dye were performed using the FDTD method, as detailed in the *Methods* section above. In the simulated model, the dye is represented as a point dipole positioned near an AuNS at a specific distance. Both the distance and the size of the sphere were adjusted to align with our experimental conditions.

Figure 6(a) displays the simulated electric field of the 25 nm AuNS in the presence of a light source (total field scatter) across a wavelength range of 400 to 900 nm. The presence of an electric field between 700 and 800 nm corresponded to the absorption tail end of the AuNS in the same region, as shown in Figure 6(a), can explain the enhanced fluorescence intensity of the NIR dye in the presence of AuNS. The electric field of the point dipole ($E_0$) is also simulated under the same conditions, but without AuNS (Figure 6(b)). Figure 6(c) shows the electric field of the 25 nm AuNS-dye complex, where the dye is positioned 6 nm away in the Z direction from the AuNS. The electric field of the AuNS-dye complex dependence on the size of the nanoparticle and its distance from the dye is shown in Figure 6(d), corroborate the curves shown in Figures 5(j) and 5(k), suggesting that nanoparticle sizes larger than 15 nm and distances shorter than ~12 nm are preferable for greater enhancement .The distances of 6 nm, 12 nm, and 30 nm were considered to PEG chains of 1 kDa, 2 kDa, and 5 kDa, respectively, as the 1 kDa PEG chain corresponds to an approximate distance of 6 nm.[31] Additionally, the quantum efficiency (Supporting Information, Figure S5) and the radiative and non-radiative losses of the dye in the presence of AuNS (Supporting Information, Figure S6) were also simulated. A high quantum efficiency for the 25 nm AuNSs at shorter distances was observed, supporting the conclusion that the electric field of AuNS in the 700-800 nm wavelength region significantly influences the utilization of AuNSs-NIR dye for fluorescence imaging in the NIR-I region. The simulation results of the scattering cross-section of AuNSs with different diameters are also provided (Supporting Information, Figure S7), indicating that the scattering cross-section increases with the diameter of the AuNSs and enhances the fluorescence intensity of larger nanoparticles conjugated to the dye compared to smaller nanoparticles.

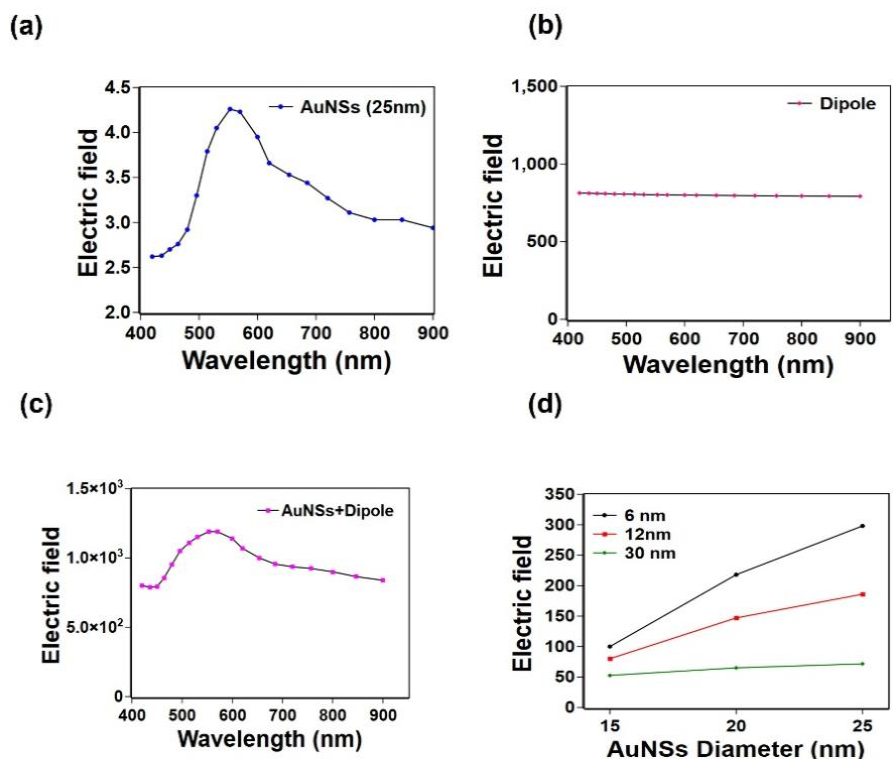

**Figure 6: FDTD simulation of the AuND model with dye as a dipole source.** (**a**) The electric field of 25 nm AuNS in the presence of total-field scatter source of wavelengths 400-900 nm. (**b**) The electric field of the dipole source. (**c**) The electric field intensity of a 25 nm AuNS+ dipole complex at 400-900 nm. (**d**) The electric field intensity of the AuNDs is calculated in FDTD at different distances at 800 nm wavelength of the dipole source. The electric field profiles and the FDTD simulation model are shown Supporting Information (Figure S8).

*Fluorescence lifetime analysis of AuNDs*

The second above-mentioned factor that might influence the dye's fluorescence intensity, modification in the dye's radiative decay rate, was addressed by measuring the FLT of synthesized AuNDs using the TCSPC method and time-gated fluorescence acquisition. Figures 7(a)-(d) show that as the length of the PEG chains increase, the FLT of AuNSs-dye also increases, particularly noticeable in the largest nanoparticles (with diameters of 20 and 25 nm). This trend in FLT is in contrast with their emission behavior. For instance, in the case of 25 nm AuNS, its FLT increases from 0.16 ns to 0.3 ns across PEG chain lengths (Figure 7(a)-(d)), while their fluorescence intensity decreases from $5.3 \times 10^6$ to $5.4 \times 10^5$ (Figure 5(i)). The decreased FLT of shorter PEGylated AuNDs can be attributed to the fluorophore being in closer proximity to the nanoparticle. This reduced distance enhances radiative decay, driven by the nanoparticle's electromagnetic field[32]. Metallic nanoparticles can enhance the local electric field near their surface, leading to increased excitation rates and a higher radiative decay rate of nearby fluorophores. In the AuNS-IRdye 800 complex, the local electric field is strongest when AuNSs are PEGylated with 1 kDa PEG and conjugated to the dye, but it decreases as the PEG chain length increases, as shown in Figure 6(d). This enhanced electric field promotes a higher radiative decay rate and reduces FLT, as demonstrated in Supplementary Information Figure S6.

Figures 7(a-d) also show that the FLT of 5 nm AuNDs remains comparable to that of the free dye. Smaller nanoparticles exhibit minimal impact on FLT modulation in the metallic dye conjugate. A study by Joron et al.[29] observed that the confined structure of fluorescent proteins enhances both QY and FLT by restricting photoisomerization, thereby reducing non-fluorescent transitions and minimizing non-radiative decay pathways. In contrast, our investigation of fluorophore conjugation with AuNSs via amide bond formation revealed enhanced fluorescence but a shorter FLTs. This behavior can be explained by the metal-enhanced fluorescence phenomenon[33]. According to the equations governing FLT and QY in the presence of metallic nanoparticles, these factors are influenced by an additional radiative decay term introduced by the metallic nanoparticle[33]. This term reduces the non-radiative decay rate, thereby enhancing the QY of the AuNDs. Additionally, when the fluorophore is conjugated to a nanoparticle, cis-trans photoisomerization is suppressed by limiting transitions between photon-emitting isomers and non-emissive intermediates. This suppression decreases non-radiative transitions, further increasing fluorescence quantum yield.

When spatial constraints are relaxed, photoisomerization becomes more feasible, leading to a reduction in fluorescence lifetime. Moreover, the amide bond between the fluorophore and AuNSs imposes spatial restrictions on the fluorophore, which enhances quantum yield by decreasing non-radiative decay rates, ultimately contributing to a shortened fluorescence lifetime.

The reduction in fluorescence lifetime can also be attributed to changes in the fluorophore's refractive index, as described by the Strickler-Berg equation. Supporting studies indicate that variations in refractive index in the presence of macromolecules, such as PEG chains, influence fluorescent proteins like mCherry, resulting in decreased fluorescence lifetime.[34]

Therefore, the reduction in fluorescence lifetime of AuNDs is a combined effect of all the mentioned factors arising from the conjugation of the dye to PEGylated AuNSs. All these effects were strongly influenced by factors like the distance between the nanoparticle and the dye, as well as the AuNS diameter, which modulates the radiative decay rate. The observed MEF phenomenon suggests that the reduction in FLT leads to enhanced fluorescence intensity and stability of the fluorescence emission spectra in different microenvironments (Supporting Information, Figure S9).making it a potential tool for imaging .[29,30,35]

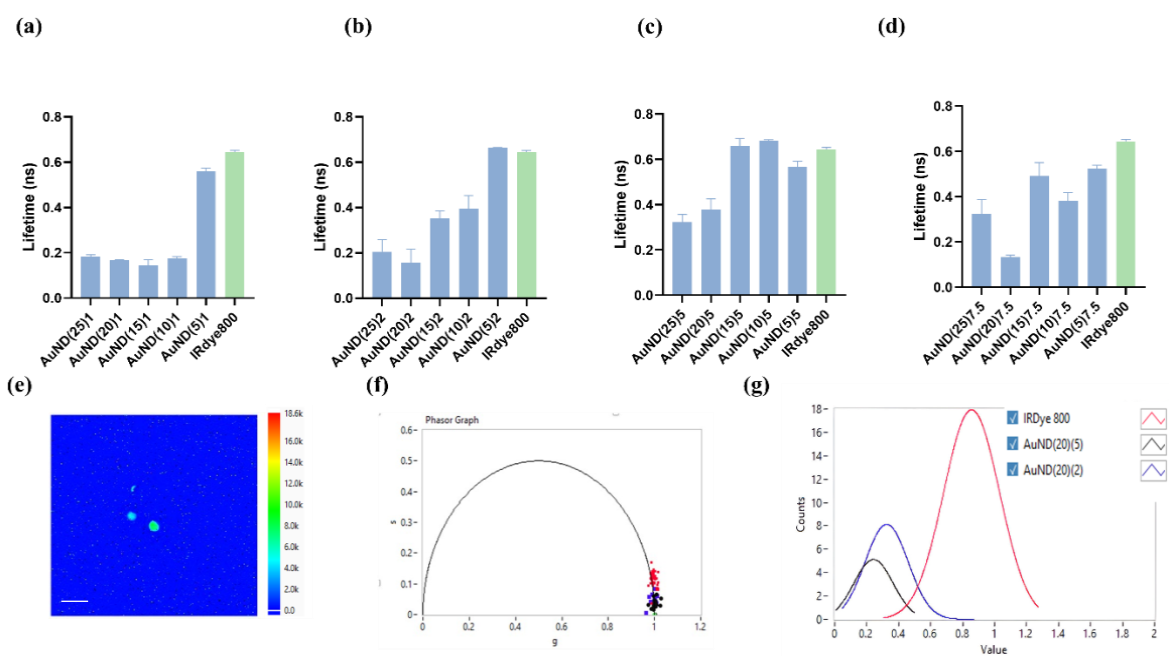

**Figure 7: FLT of the AuNDs.** FLT of different diameter AuNSs conjugated to the dye via **(a)** 1 kDa **(b)** 2 kDa **(c)** 5 kDa and **(d)** 7.5 kDa PEG chains, respectively. **(e)** Intensity image displaying the fluorescence from the samples; IRdye 800, AuND(20)2 and AuND(20)5. Scale bar: 7 mm. **(f)** Phasor scatter plots for the three samples shown in (e). Phasors were calibrated using the IRF lifetime ($\tau$ = 0 ns) and a phasor harmonic frequency f = 20 MHz. **(g)** Phase lifetime histograms corresponding to the phasor plots in (f). The histogram was Gaussian-fitted to extract a peak phase lifetime. The ROIs varied for the three samples, thus the phasor count values differ in the graphs. Background noise has been eliminated using consistent dark (no illumination) gated images.

Having characterized the lifetime behaviour of the IRdye 800 and the AuNDs, the samples were imaged using a highly sensitive time-gated SPAD camera (Figure 7 (e-g)). The main objective of these experiments was to explore multiplexing FLI, using a single dye when conjugated to the AuNS at varying AuNS-dye distances. A schematic illustration of the wide-field FLI setup utilized in our investigations can be found in our previous work.[36] The fluorescence image of three representative dyes, IRdye 800, AuND(20)2, and AuND(20)5, as captured by the SPAD512S, is shown in Figure 7(e). Phasor analyses of intensity spots in Figure 7(e) are displayed in Figure 7(f), allowing for the differentiation of the samples within a single image frame, based on their fluorescence lifetimes. This visual representation serves as an effective means to distinguish between the various samples, despite the high proximity of the different FLT values. Histograms detailing the number of phasor counts for each FLT were constructed from these phasor plots (Figure 7(g)). The phase lifetimes were computed for each pixel within a region of interest (ROI) in each intensity spot (ROIs were not the same for each spot), furnishing data essential for constructing histograms that represent the count distribution for each phase lifetime. The mean of the histogram was used to calculate phase lifetimes for each individual dye. The resulting fluorescence lifetimes for different samples partly align with the TCSPC results illustrated in Figure 7(c) and 7(b), suggesting the following fluorescence lifetimes: 0.9 ns for IRdye 800, 0.26 ns, and 0.35 ns for AuND(20)5 and AuND(20)2, respectively. While the IRdye 800 showed a phasor-based FLT different from the one that was achieved using the TCSPC set up, the AuNDs showed results very similar to the TCSPC results.

*Steric stabilization of the IRdye 800 NHS*

The third factor that might influence the fluorescence enhancement effect, is the steric stabilization of the cyanine NIR dye when conjugated to nanoparticles.[37-39] The subsequent paragraphs present emission results for IRdye 800 when conjugated to ZnO nanoparticles. These experiments aimed to ascertain whether nanoparticles lacking SPR still induce a change in the dye's emission, attributable solely to chemical stabilization. Figure 8(a) presents the absorption and TEM image of synthesized ZnO nanoparticles, resulting with an absorption at 370 nm and an average particle size of 20 nm. Figure 8(b) shows fluorescence emission spectra ZnO conjugated to IRdye 800.

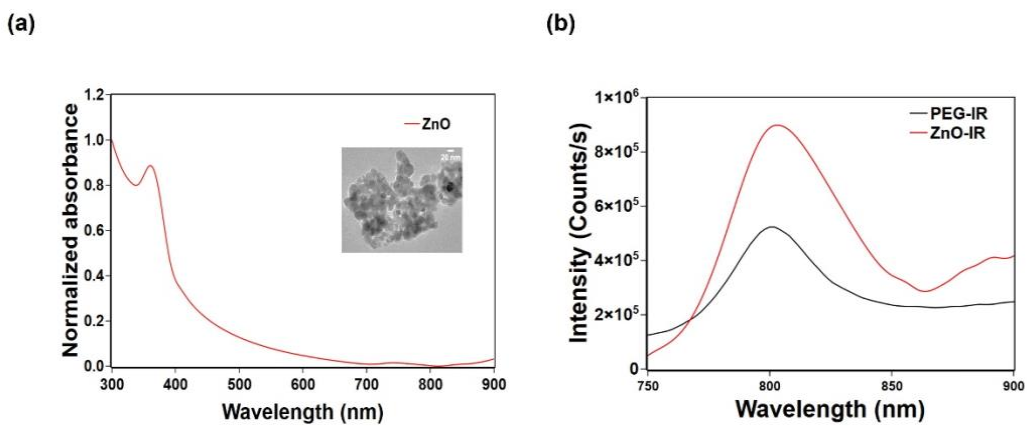

**Figure 8: The impact of steric stabilization on the emission of IRdye 800 using ZnO nanoparticles. (a)** UV-VIS absorption spectrum of the ZnO (370 nm) Inset: TEM image of the ZnO. The scale bar measures 20 nm. **(b)** The fluorescence emission spectra of PEG-IRdye 800 and ZnO-IR with an emission peak at 807 nm and 803 nm, respectively.

The AuNS and ZnO were conjugated to the dye using the 1 kDa PEG chain as a space layer and linker. The fluorescence intensity of the ZnO-IR complex was higher compared to that of the PEG-IRdye 800 complex. This increased fluorescence can be explained by a reduction in photoisomerization, which is attributed to the steric hindrance of the ZnO-IR complex compared to the free dye. [37-39] The formation of an amide bond between the PEGylated nanoparticle and the dye introduced steric hindrance at the C–C bond of the dye, resulting in increased fluorescence intensity compared to the unbound dye.[40,41] The PEG chains, adsorbed on the surface of the ZnO nanoparticles,[42] facilitate the conjugation of the dye to the -NH$_2$ group on the PEG chain causing the higher fluorescence of the dye in the presence of ZnO. The improved fluorescence of IRdye 800 in the presence of AuNS is thus a combined phenomenon of the MEF effect and also the steric stabilization of the dye arises from conjugation.

**Discussion**

Fluorescence imaging is one of the most widely utilized non-invasive and non-destructive techniques for visualizing biological processes.[43] This method offers rapid imaging with high sensitivity, producing images within milliseconds and achieving resolutions up to tens of nanometers.[44] However, the majority of fluorescence-based biomolecular imaging techniques rely on the visible spectrum (400–750 nm), which is suboptimal for *in vivo* imaging due to high autofluorescence and limited tissue penetration at these wavelengths, resulting in shallow imaging depths and reduced sensitivity and spatial resolution.[45] A potential solution to this challenge is to utilize NIR-I radiation, which offers deeper tissue penetration and enhanced imaging resolution compared to visible wavelengths.[46] Augmenting the fluorescence of commercially accessible NIR dyes holds paramount significance in the realm of biological imaging, owing to the prevalent challenges of low intensity and inadequate photostability encountered particularly in *in vivo* applications.[47] The conjugation of inert and biocompatible metals like gold can improve the intensity and photostability of the dye.[10,48]

In this work, we investigated the dependence on the size/diameter of gold nanospheres on the photophysical properties of IRdye 800 along with the distance of the fluorophore from the nanoparticle. Gold nanorods (AuNRs) and nano stars, which exhibit absorption in the NIR-I region, have been widely explored in NIR imaging. Thus, e.g., studies by Fixler and colleagues demonstrated the feasibility of conjugating AuNRs with fluorophores via a polyethylene glycol (PEG) linker, enabling fluorescence lifetime imaging (FLIM) in the NIR-I region[49]. Still, AuNRs have lower uptake by mammalian cells [8,50] and exhibit potential higher toxicity compared to AuNSs[51]. Emelianov and his team highlighted that aggregated AuNSs enhance NIR fluorescence[52]. In our work, rather than relying on aggregation, we focus on conjugating AuNSs with NIR fluorophores. This approach significantly enhances the fluorescence intensity of the fluorophore, offering improved fluorescence imaging capabilities.

Figure 5 clearly demonstrates that optimal fluorescence enhancement is achieved with a shorter PEG chain distance and a larger nanoparticle size. The MEF phenomenon partially depends on the scattering cross-section of the metallic nanoparticles.[53] The fluorescence enhancement of dye conjugated to larger AuNSs are due to the larger scattering cross-section and absorption cross-section of particles compared to small-sized nanoparticles.[54] From our studies as the diameter of the spheres increased the fluorescence intensity also increased in the case of the shorter distance. The fluorescence enhancement in MEF is related to the plasmon-coupled resonance energy transfer (PC-RET) process. Recent studies showed that in the PC-RET process, it is not necessary for spectral overlap of the absorption of the donor and the emission of the acceptor.[55] Alternatively, a notable plasmonic effect may occur at a distinct wavelength compared to the extinction peak of the plasmonic structure. This factor influenced the fluorescence enhancement of the dye as it was conjugated to AuNS which does not have any spectral overlap with dye. On the other hand, when fluorophores are near plasmonic nanoparticles, some of their excited state energies are transferred to induced surface plasmons. This leads to two notable effects: first, there is an increase in fluorescence emission from the combined metal-fluorophore system while maintaining the spectral properties of the fluorophores. Second, there is a decrease in the excited-state decay time (fluorescence lifetime), which enhances the photostability of the fluorophores by reducing the time they spend in a reactive excited state, causing enhanced fluorescence.[56]

In the simulation study of the metallic dye system, we observed that the enhancement of fluorescence is strongly influenced by the local electric fields generated by the interaction between the metallic nanoparticles and the dye molecules. As the size of the nanoparticles increased, the LSPR intensified, resulting in stronger electric fields around the nanoparticle-dye complex compared to the free dye. This led to an increased excitation rate of the dye and thus to an increased fluorescence intensity.[57] This size-dependent amplification suggests that nanoparticle size control is critical for optimizing fluorescence intensity in plasmonic dye systems along with shorter distance.

In addition, PEGylated nanoparticles, when conjugated to a dye, dramatically lower the efficiency of photoisomerization, leading to fluorescence enhancement of the cyanine dyes.[48,58] This property of the cyanine dyes leads to the fluorescence enhancement of the dye when it is conjugated to AuNS and ZnO. Thus, the fluorescence enhancement of IRdye 800 can be achieved through AuNS of larger diameter (15-25 nm) opens a new window for the utilization of AuNS in NIR window.

**Conclusion**

The use of AuNSs in biological imaging offers significant advantages, including targeted delivery and imaging, low cytotoxicity and efficient cellular uptake. However, their application has been largely limited to the visible spectrum due to their strong absorption in this region. In this study, we demonstrated the potential of AuNSs for NIR-I imaging by conjugating them with a NIR fluorophore. Our findings indicate that this approach overcomes the limitations of conventional NIR-I fluorophores, which often suffer from low fluorescence intensity and unstable fluorescence lifetime, thereby enhancing their stability and efficiency for imaging applications. Additionally, we identified three key mechanisms responsible for fluorescence enhancement, providing a framework for fine-tuning AuNS-based NIR-I imaging.

Our results show that PEGylated AuNSs with diameters of 15 nm 20 nm and 25 nm offer optimal fluorescence enhancement, with a fluorophore-nanoparticle separation distance of 6–12 nm being the most effective. Our results also demonstrate that 15 nm AuNSs, when coated with a 1 kDa PEG chain, show enhanced fluorescence compared to free NIR fluorophore. We found that the minimal limit for AuNSs application in NIR imaging is 15 nm, while AuNSs below this size did not provide fluorescence enhancement for effective imaging.

The integration of AuNSs in fluorescence imaging paves the way for dual-modal imaging applications. Future research could focus on combining computed tomography (CT), which extensively employs AuNSs for structural imaging, with fluorescence lifetime imaging for functional information, to create a complementary imaging modality with high spatial resolution and sensitivity. While our in vitro studies demonstrate the promise of AuNS-based fluorescent NIR nanoprobes[16], further in vivo investigations are essential to fully explore their clinical applications.

**Acknowledgment:** This work was supported by the Council for Higher Education, Israel (No. RA2100000248).

**Supporting Information**

The Supporting Information is available free of charge and contains additional characterization of the AuNS, as well as additional results for the electromagnetic field of the AuNDs.

**Author contributions**

RA conceived and designed the study and revised the manuscript. NC designed and implemented most experiments and wrote manuscripts and MM helped the PEGylation of AuNSs.

**Declarations**

**Competing interests** The authors declare no competing interests.

**Consent for publication** All authors agree to the publication of this manuscript.

**Data availability**

The data that support the findings of this study are available from the corresponding authors upon reasonable request.